\documentclass[a4paper,fleqn]{cas-sc}

\usepackage[numbers,sort&compress]{natbib}
\usepackage{amsmath,amssymb,amsthm}
\usepackage{booktabs,tabularx}
\usepackage{float}
\usepackage{tikz}
\usepackage[capitalise,noabbrev]{cleveref}
\usepackage{todonotes}
\newfloat{algorithm}{tbp}{loa}
\floatname{algorithm}{Algorithm}

\newtheorem{thm}{Theorem}[section]
\newtheorem{lem}[thm]{Lemma}
\newtheorem{fact}[thm]{Fact}

\newtheorem{cor}[thm]{Corollary}
\newtheorem{prop}[thm]{Proposition}
\theoremstyle{remark}
\newtheorem{rem}[thm]{Remark}
\theoremstyle{plain}
\newcommand{\defn}[1]{\emph{#1}}
\newcommand{\cd}{\textsc{Cluster Deletion}}
\newcommand{\cp}{\textsc{Clique Partition}}
\newcommand{\mc}{\textsc{Max Clique}}
\newcommand{\occp}{\textsc{Optimum Cost Chromatic Partition}}
\newcommand{\greedy}{{\textsc{Greedy}}}

\newcommand{\edmonds}{{\textsc{Greedy Edmonds}}}
\newcommand{\opt}{{\mathsf{OPT}}}
\newcommand{\cost}{{\mathsf{cost}}}
\newcommand{\score}{{\mathsf{sc}}}

\begin{document}

\let\WriteBookmarks\relax
\def\floatpagepagefraction{1}
\def\textpagefraction{.001}

\shorttitle{Cluster deletion and clique partitioning}
\shortauthors{Galesi, Huynh, Patey and Ranjbar}

\title[mode=title]{Cluster deletion in cographs, permutation graphs, and graphs with bounded clique number}

\author[1]{Nicola Galesi}[orcid=0000-0002-8522-362X]
\ead{galesi@diag.uniroma1.it}
\ead[url]{https://sites.google.com/diag.uniroma1.it/nicolagalesi}

\author[2]{Tony Huynh}[orcid=0000-0002-6908-923X]
\cormark[1]
\ead{tony@ibs.re.kr}
\ead[url]{https://sites.google.com/site/matroidintersection/home/}

\author[3]{Arnaud Patey}[]
\ead{arnaud.patey@ens-rennes.fr}

\author[4]{Fariba Ranjbar}[orcid=0000-0001-6432-3683]
\ead{fariba.ranjbar@luiss.it}

\affiliation[1]{organization={Department of Computer, Control and Management Engineering,
Sapienza University of Rome},
addressline={Via Ariosto 25},
city={Rome},
postcode={00185},
country={Italy}}

\affiliation[2]{organization={Discrete Mathematics Group, Institute for Basic Science},
addressline={55 Expo-ro, Yuseong-gu},
city={Daejeon},
postcode={34126},
country={Republic of Korea}}

\affiliation[3]{organization={ENS Rennes},
addressline={11 Avenue Robert Schuman},
city={Bruz},
postcode={35170},
country={France}}

\affiliation[4]{organization={Department of AI, Data and Decision Sciences,
Luiss Guido Carli},
addressline={Viale Romania 32},
city={Rome},
postcode={00197},
country={Italy}}

\cortext[1]{Corresponding author}

\begin{abstract}
The \cd{} problem asks for a minimum-size edge set whose deletion turns
a graph into a disjoint union of complete graphs. Equivalently, the
\cp{} problem asks for a partition of the vertex set into cliques that
maximizes the number of edges within the parts. We give a simpler proof of a result of Gao, Hare, and Nastos~\cite{GHN13},
that \cd{} is polynomial-time solvable on cographs. In addition, we show that the natural linear programming formulation of \cp{} is exact on cographs.

We then show that \cd{} is NP-complete on permutation graphs, which are a superclass of cographs.  This answers an open question of Konstantinidis and Papadopoulos~\cite{konstantinidis2021cluster}. We also exhibit a permutation graph on nine vertices for which the linear programming formulation is not exact.


Finally, for graphs
with clique number at most $c$, we give a polynomial-time
$2\binom{c}{2}/(\binom{c}{2}+1)$-approximation algorithm for \cp{}.
More generally, the algorithm runs in polynomial time on every graph
class for which a maximum clique can be found in polynomial time. For
each fixed $c\geq 3$, we also construct infinitely many examples attaining
the stated approximation ratio.  The same examples show that, for \cd{}, the algorithm is a $2$-approximation and no better, for every fixed $c \geq 3$.
\end{abstract}



\maketitle

\section{Introduction}

Graph clustering is a fundamental problem in graph theory with numerous
real-world applications, including the identification of functionally
related genes in computational biology \cite{Ben1999}, community
detection in social networks \cite{FORTUNATO2010}, image segmentation
\cite{Shi2000}, and machine learning \cite{bansal2004,panli2020}. The
goal is to partition the vertices into \defn{clusters} with many edges
within clusters and few edges between clusters. Since a complete graph
is the densest possible graph, this motivates the following approach.

\medskip

\noindent\fbox{\parbox{\dimexpr\linewidth-2\fboxsep-2\fboxrule-4pt\relax}{%
$\cd$:
\\
{\em Input:} A graph $G$.\\
{\em Output:} A minimum size set of edges $X \subseteq E(G)$ such that $G-X$ is a disjoint union of complete graphs.} 
}

 \medskip
 

A disjoint union of complete graphs is also called a \defn{cluster graph}. Given a graph $G$ as input, the \cd{} problem is equivalent to maximizing the number of edges in a spanning subgraph $H$ of $G$ which is a cluster graph.  We call the equivalent `complementary' version the $\cp{}$ problem.
\medskip

\noindent\fbox{\parbox{\dimexpr\linewidth-2\fboxsep-2\fboxrule-4pt\relax}{%
$\cp$:
\\
{\em Input:} A graph $G$.\\
{\em Output:} A partition $(X_1, \dots, X_s)$ of $V(G)$ such that each $X_i$ is a clique of $G$ and $\sum_{i=1}^s |E(X_i)|$ is maximum.}
}

 \medskip

\noindent Whenever we speak of NP-completeness, we refer to the natural decision versions of these problems. Since $\cd{}$ has a solution of size at most $k$ if and only if $\cp{}$ has a solution of size at least $|E(G)|-k$, the two decision problems are equivalent, and both are clearly in $\mathsf{NP}$.

\medskip

\noindent \textbf{Previous work.} Shamir et al.~\cite{SHAMIR2004} proved that \cd{} is NP-hard for general graphs.  Despite its computational difficulty, the problem has been extensively explored within the realms of parameterized complexity \cite{gramm2005,damaschke2009,BOCKER2011,GHN13,Bathie2022,italiano2025}, approximation algorithms~\cite{CHARIKAR2005,DESSMARK2007,Puleo2015,veldt2018,veldt2022,balmaseda2024}, and restricted graph classes.  

For instance, \cd{} is polynomial time solvable on cographs \cite{GHN13}, split graphs \cite{BONOMO2015,konstantinidis2021cluster}, (proper) interval graphs \cite{BONOMO2015,konstantinidis2021cluster}, $P_4$- and paw-free graphs \cite{komus2018} and graphs with maximum degree at most three \cite{KOMUSIEWICZ2012}. On the other hand, NP-hardness persists on $C_4$-free graphs with maximum degree four \cite{KOMUSIEWICZ2012}, planar graphs \cite{golovach2018}, chordal graphs \cite{BONOMO2015}, $C_5$-free graphs and (2$K_2$, 3$K_1$)-free graphs \cite{GHN13}. 

Dessmark et al.~\cite{DESSMARK2007} introduced a simple greedy strategy (which we call \greedy)
of repeatedly choosing maximum cliques, and proved that \greedy{} is a 2-approximation for both \cd{} and \cp{}.  However, \greedy{} does not necessarily run in polynomial time, since it requires \mc{} as a subroutine. Gao et al.~\cite{GHN13} proved that \greedy{} optimally solves \cp{} (and hence also \cd) on cographs in polynomial time.  

Interestingly, the weighted version of \cd{} often exhibits greater computational hardness even on instances where the unweighted version is tractable. For instance, while unweighted \cd{} is tractable on split graphs and cographs, the weighted version remains NP-hard on these classes \cite{BONOMO2015}. 
There are also other popular generalizations of \cd{}, including \emph{correlation clustering~}\cite{bansal2004,SHAMIR2004}, where one is allowed to delete or add edges.  In this work, we focus on the unweighted variant of \cd{}.

From the perspective of approximation algorithms, there have been a series of results for \cd{} with progressively improved guarantees. The first approximation algorithm for \cd{} achieved an approximation ratio of 4 using a linear programming (LP) relaxation \cite{CHARIKAR2005}.  Recent work of Balmaseda et al.~\cite{balmaseda2024} refined the analysis of two earlier algorithms, improving the approximation ratio from 4 to 3.  The state of the art is the $2$-approximation of Veldt et al.~\cite{veldt2018}. However, the LP-based approaches are computationally expensive due to the large number of constraints involved. To address this, recent work has introduced faster algorithms by leveraging alternative LP formulations such as Strong Triadic Closure (STC) labeling \cite{sintos2014}, resulting in practical $4$-approximations \cite{veldt2022}. Veldt~\cite{veldt2022} also proposed the first combinatorial algorithm, \defn{MatchFlipPivot}, which achieves similar approximation guarantees at significantly reduced runtime. 

The approximation status of the \emph{vertex} version of \cd{}, in
which vertices rather than edges are deleted, was settled by Aprile
et al.~\cite{ADFH23}. They gave a $2$-approximation algorithm, which
is best possible assuming the \defn{Unique Games Conjecture}.

\noindent \textbf{Our contributions.}
We begin by giving a much simpler proof of correctness of the main result from~\cite{GHN13}.
\begin{thm}[\cite{GHN13}] \label{thm:cograph}
    For every cograph $G$, \greedy{} outputs an optimal clique partition of $G$. 
\end{thm}

The proof in~\cite{GHN13} uses a quite subtle induction, which requires carefully extending a given clique partition while controlling the number of parts of the clique partition.  We instead use the well-known fact that every cograph is either the disjoint union of two smaller cographs or the join of two smaller cographs.  This gives a very clean inductive proof of~\cref{thm:cograph}. We also note that Theorem 4 from~\cite{GHN13} is a direct consequence of the majorization inequality (Karamata's inequality~\cite{MOA11}). Our method also proves that the natural linear programming formulation is exact on cographs.  To be precise, let $\mathcal{C}(G)$ be the family of all
nonempty cliques of $G$, and consider
\begin{equation*}
\begin{aligned}
\max\quad&
   \sum_{K\in\mathcal{C}(G)}\binom{|K|}{2}x_K\\
\text{subject to}\quad&
   \sum_{\substack{K\in\mathcal{C}(G)\\v\in K}}x_K=1
      &&(v\in V(G)),\\
&x_K\geq 0 &&(K\in\mathcal{C}(G)).
\end{aligned}
\tag{LP}\label{lp:clique-partition}
\end{equation*}

\begin{thm} \label{thm:cograph-lp}
For every cograph $G$, \eqref{lp:clique-partition} has an optimal
solution in which every variable is in $\{0,1\}$.
\end{thm}

We next analyze \cp{} and \cd{} on \defn{permutation graphs}, which are a superclass of cographs. We prove that both problems are NP-complete on permutation graphs.

\begin{thm} \label{thm:NPhardness}
    \cd{} and \cp{} are NP-complete on permutation graphs.
\end{thm}

\cref{thm:NPhardness} settles an open question of Konstantinidis and Papadopoulos~\cite{konstantinidis2021cluster}, and suggests that cographs are at the boundary of tractability and intractability for \cd{}. The proof is a polynomial reduction from the \occp{} problem, which Jansen~\cite{JANSEN200054} proved to be NP-complete on permutation graphs.  Since permutation graphs are exactly the graphs that are both comparability and cocomparability graphs, \cref{thm:NPhardness} also implies that \cd{} is NP-complete on comparability graphs and on cocomparability graphs (\cref{cor:comparability}). 

\Cref{thm:NPhardness} also has consequences for the linear program \eqref{lp:clique-partition}.  Although \eqref{lp:clique-partition} has exponentially many variables, it can be solved in polynomial time on permutation graphs via the ellipsoid method, so \cref{thm:NPhardness} implies that \cref{thm:cograph-lp} does not extend to permutation graphs unless $\mathsf{P}=\mathsf{NP}$.  In fact, we give an explicit permutation graph on nine vertices for which \eqref{lp:clique-partition} is not exact (\cref{prop:lp-perm}).


 Finally, we introduce another variant of \greedy{}.  For a graph $G$, we let $\omega(G)$ and $\omega'(G)$ be the number of vertices and edges, respectively, in a maximum clique of $G$.  The main idea is very simple: if $\omega(G) \leq 2$, then a maximum clique partition is just a maximum matching, and hence can be found in polynomial-time via Edmonds' matching algorithm \cite{Edmonds_1965}.  Therefore, we propose the following modification of \greedy.  We run \greedy{} as long as $\omega(G) \geq 3$.  Once $\omega(G) \leq 2$, we instead use Edmonds' matching algorithm on the remaining graph and add the maximum matching to the clique partition.  We call this algorithm \edmonds{}. We prove that for every graph $G$, \edmonds{} is a $\frac{2\omega'(G)}{\omega'(G)+1}$-approximation algorithm for \cp.

 \begin{thm} \label{thm:main-summary}
    For every graph $G$, \edmonds{} returns a clique partition $\mathcal{X}$ such that \[\opt \leq \frac{2\omega'(G)}{\omega'(G)+1}|E(\mathcal{X})|,\] where $\opt$ is the number of edges in an optimal clique partition of $G$. 
\end{thm}


Although optimal solutions to \cp{} correspond to optimal solutions to \cd, it is not necessarily true that an $\alpha$-approximate solution to \cp{} corresponds to an $\alpha$-approximate solution to \cd. Indeed, for general $n$-vertex graphs, Dessmark et al.~\cite{DESSMARK2007} proved that there is no polynomial time $n^{1-O(\frac{1}{(\log n)^\gamma})}$-approximation algorithm for \cp{} (for some constant $\gamma$), unless $\mathsf{NP} \subseteq \mathsf{ZPTIME}(2^{(\log n)^{O(1)}})$.  On the other hand, as already mentioned, there are constant factor approximations for \cd.  

Note that \edmonds{} runs in polynomial time on any graph class for which a maximum clique can be found in polynomial time.  Examples of such graph classes include graphs of bounded treewidth, perfect graphs (which include interval graphs and chordal graphs), distance-hereditary graphs, and graphs with bounded clique number.
To our knowledge, for graphs of bounded clique number, \edmonds{} is
the first polynomial-time algorithm that achieves an approximation
ratio better than $2$ for \cp. For general graphs $G$, the best
approximation algorithm for \cp{} is the $\omega(G)$-approximation
algorithm from~\cite{DESSMARK2007}.

In view of \Cref{thm:NPhardness}, it is natural to ask for approximation algorithms for \cp{} on permutation graphs.  Since maximum cliques and maximum matchings can be computed efficiently in permutation graphs, \Cref{thm:main-summary} yields the following corollary.

\begin{cor} \label{cor:perm-approx}
  Given an $n$-vertex permutation graph $G$, \edmonds{} returns in
  $O(n^2\log n)$ time a clique partition $\mathcal{X}$ such that
  \[
  \opt \leq
  \frac{2\omega'(G)}{\omega'(G)+1}|E(\mathcal{X})|,
  \]
  where $\opt$ is the number of edges in an optimal clique partition
  of $G$.
\end{cor}

\begin{proof}
    Let $G$ be an $n$-vertex permutation graph.  A permutation representation of $G$ can be found in $O(n+|E(G)|)$ time~\cite{MS99}. Since a maximum clique in a permutation graph can be found in $O(n \log n)$ time from such a representation~\cite{Fredman75}, the greedy portion of \edmonds{} can be made to run in $O(n^2 \log n)$ time.
    Finally, Rhee and Liang~\cite{RL95} proved that a maximum matching of an $n$-vertex permutation graph can be found in $O(n \log \log n)$-time. Thus, the total running time is $O(n^2\log n)$.
\end{proof}

Finally, we show that the approximation ratio in~\Cref{thm:main-summary} is best possible for \edmonds{}, for every fixed value of the clique number, and that the same examples show that for \cd{} the approximation ratio of \edmonds{} is exactly $2$.

\begin{thm} \label{thm:tight-summary}
        For each fixed $\ell \geq 3$, there are infinitely many graphs $G$ with $\omega(G)=\ell$ and clique partitions $\mathcal{X}$ that can be output by \edmonds{} such that
        \[
        \opt=
        \frac{2\binom{\ell}{2}}{\binom{\ell}{2}+1}
        |E(\mathcal{X})|
        \qquad\text{and}\qquad
        |E(\overline{\mathcal{X}})|=2\,\opt_{\mathrm{CD}},
        \]
        where $\opt$ is the maximum number of edges in a clique partition of $G$ and $\opt_{\mathrm{CD}}=|E(G)|-\opt$ is the minimum number of edges whose deletion turns $G$ into a cluster graph.
   \end{thm}
 
\section{Preliminaries}

All graphs in this paper are assumed to be simple and undirected. We use standard graph theoretical notation and terminology (see~\cite{diestel25}). For example, we say that a graph $G$ is \defn{$H$-free}, if it does not contain $H$ as an induced subgraph. Note that $G$ is a cluster graph if and only if $G$ is $P_3$-free, where $P_3$ is a 3-vertex path.  A graph $G$ is a \defn{cograph} if
it is $P_4$-free.  The \defn{complement} of a graph $G$ is denoted by $\overline{G}$.  The \defn{disjoint union} of two graphs $G_1$ and $G_2$ is denoted by $G_1 \sqcup G_2$, and the \defn{join} $G_1 \vee G_2$ is the graph obtained from $G_1 \sqcup G_2$ by adding all edges between $V(G_1)$ and $V(G_2)$.

A \defn{clique} in a graph $G$ is a set of pairwise adjacent vertices of $G$, and a \defn{stable set} is a set of pairwise nonadjacent vertices.  Thus, the stable sets of $G$ are exactly the cliques of $\overline{G}$.  A \defn{clique partition} of $G$ is a partition of $V(G)$ into cliques of $G$.  If $\mathcal{X}$ is a clique partition of $G$, we let $E(\mathcal{X})$ and $E(\overline{\mathcal{X}})$ denote the set of edges of $G$ contained (respectively, not contained) in the cliques of $\mathcal{X}$.

For each $n \in \mathbb{N}$, we let $[n]:=\{1, \dots, n\}$ and $S_n$ denote the set of all permutations of $[n]$. We write a permutation as a word
 $\pi = [\pi_1, \dots, \pi_n]$, and write $\operatorname{pos}_\pi(v)$ for the position of the value $v$ in this word. The \defn{permutation graph} associated with $\pi$, denoted by $G_\pi$, has vertex set $[n]$ and edge set
\[
E(G_\pi) := \left\{ \{u,v\} : u<v \text{ and } \operatorname{pos}_\pi(u)>\operatorname{pos}_\pi(v) \right\}.
\]
Equivalently, if $i<j$ and $\pi_i>\pi_j$, then $\{\pi_i,\pi_j\}$ is an edge. Thus, cliques of $G_\pi$ are precisely the decreasing subsequences of the word $\pi$, and stable sets of $G_\pi$ are precisely the increasing subsequences of $\pi$.  In particular, if $\pi^{\mathrm{R}}:=[\pi_n, \dots, \pi_1]$ is the reverse of $\pi$, then $G_{\pi^{\mathrm{R}}}=\overline{G_\pi}$, so the class of permutation graphs is closed under complementation.  A graph is a permutation graph if and only if it is isomorphic to $G_\pi$ for some permutation $\pi$. Given a permutation graph, such a permutation can be found in linear time~\cite{MS99}. Permutation graphs were originally introduced by \cite{EPL1972} and \cite{PLE1971}. A comprehensive overview of their structural and algorithmic properties is provided in \cite{golumbic2004}.

Let $\alpha \geq 1$.  An \defn{$\alpha$-approximation algorithm} is an algorithm which returns a solution which is within a multiplicative factor of $\alpha$ to an optimal solution.  

\section{A Greedy Algorithm}
In this section we review the algorithm for $\cp$ introduced by Dessmark et al.~\cite{DESSMARK2007}.
The algorithm requires $\mc$ as a subroutine, and therefore does not run in polynomial time for general graphs.  However, in every graph class for which $\mc$ can be computed in polynomial time, the algorithm runs in polynomial time. We call the algorithm \greedy, since it proceeds greedily.  The description of \greedy{} is as follows.  Let $G$ be a graph and $X$ be a maximum clique of $G$.  \greedy{} adds $X$ to the clique partition and then recurses on $G-X$.

\begin{algorithm}[h!]
\caption{\greedy}
\begin{tabularx}{\columnwidth}{@{}rX@{}}
\textbf{Input:} & A graph $G$.\\
\textbf{Output:} & A clique partition $\mathcal X$ of $G$.\\[2pt]
1. & Set $\mathcal X=\emptyset$.\\
2. & While $V(G)\neq\emptyset$:\\
   & \quad choose a maximum clique $X$ of $G$;\\
   & \quad set $\mathcal X\gets\mathcal X\cup\{X\}$ and $G\gets G-X$.\\
3. & Return $\mathcal X$.
\end{tabularx}

\end{algorithm}

  
Dessmark et al.~\cite{DESSMARK2007} proved that \greedy{} is a 2-approximation algorithm for both \cd{} and \cp.
 
\begin{thm} \label{thm:greedy1}
    For every graph $G$, \greedy{} returns a clique partition $\mathcal{X}$ such that $2|E(\mathcal{X})| \geq \opt$, where $\opt$ is the number of edges in an optimal clique partition of $G$. 
\end{thm}

\begin{thm} \label{thm:greedy2}
    For every graph $G$, \greedy{} returns a clique partition $\mathcal{X}$ such that $|E(\overline{\mathcal{X}})| \leq 2\opt$, where $\opt$ is the number of edges in an optimal solution to \cd, and $E(\overline{\mathcal{X}})$ are the edges in $G$ which are not contained in a clique of $\mathcal X$. 
\end{thm}

For completeness, we include proofs of both~\Cref{thm:greedy1} and~\Cref{thm:greedy2} in~\Cref{app:greedy}.

\section{Cluster deletion in cographs}
In this section, we give a simpler proof that \greedy{} computes an optimal clique partition in cographs, which was first proved in~\cite{GHN13}. We require the following well-known property of cographs.

\begin{lem} \label{lem:cographpartition}
    Every cograph $G$ with at least two vertices has a partition $(A,B)$ of $V(G)$ such that either $G=G[A] \sqcup G[B]$ or $G=G[A] \vee G[B]$.  
\end{lem}

For a clique partition $\mathcal{P}$, let
$p_1\geq p_2\geq\cdots$ be its clique sizes in nonincreasing order, padded by zeros. We say that $\mathcal{P}$ \defn{majorizes} a clique partition $\mathcal{Q}$, whose padded size sequence is $q_1\geq q_2\geq\cdots$, if
\[
   \sum_{i=1}^j p_i\geq \sum_{i=1}^j q_i
   \qquad\text{for every }j\geq 1.
\]

\begin{lem} \label{lem:partialorder}
If $\mathcal{P}$ majorizes $\mathcal{Q}$, then
$|E(\mathcal{P})|\geq |E(\mathcal{Q})|$.
\end{lem}

\begin{proof}
Both size sequences have the same sum, namely $|V(G)|$. The function
$x\mapsto \binom{x}{2}$ is convex on the nonnegative reals, so the
majorization inequality (equivalently, Karamata's inequality~\cite{MOA11}) gives
\[
   |E(\mathcal{P})|=\sum_i\binom{p_i}{2}
   \geq \sum_i\binom{q_i}{2}=|E(\mathcal{Q})|.
\]
\end{proof}

We prove the following statement, which is stronger than optimality.

\begin{thm} \label{thm:simpercograph}
Let $G$ be a cograph. Every clique partition of $G$ output by
\greedy{} majorizes every clique partition of $G$.
\end{thm}

\begin{proof}
We proceed by induction on $|V(G)|$. The claim is immediate when
$|V(G)|\leq 1$. Let $\mathcal{X}$ be a partition output by \greedy{},
and let $\mathcal{Y}$ be any clique partition of $G$. By
\Cref{lem:cographpartition}, there is a nontrivial partition $(A,B)$
of $V(G)$ such that $G=G[A]\sqcup G[B]$ or
$G=G[A]\vee G[B]$.

First suppose that $G=G[A]\sqcup G[B]$. Every clique lies wholly in
$A$ or wholly in $B$. Moreover, the restriction of the greedy run to
each side is clearly a greedy run on that side. Thus, the
size sequence of $\mathcal{X}$ is the nonincreasing merge of greedy
size sequences $p^A$ and $p^B$ for $G[A]$ and $G[B]$. Likewise, the
size sequence of $\mathcal{Y}$ is the nonincreasing merge of size
sequences $q^A$ and $q^B$ for clique partitions of the two sides. By
induction, $p^A$ majorizes $q^A$ and $p^B$ majorizes $q^B$. For each
$j\geq 1$,
\begin{align*}
\sum_{i=1}^j |Y_i|
 &=\max_{r+s=j}\left(\sum_{i=1}^r q^A_i+
                         \sum_{i=1}^s q^B_i\right)\\
 &\leq\max_{r+s=j}\left(\sum_{i=1}^r p^A_i+
                         \sum_{i=1}^s p^B_i\right)
  =\sum_{i=1}^j |X_i|,
\end{align*}
where the parts of $\mathcal{X}$ and $\mathcal{Y}$ are ordered by
nonincreasing size.

Now suppose that $G=G[A]\vee G[B]$. A maximum clique in every induced
subgraph of this join is the union of a maximum clique on each
nonempty side. Consequently, there are greedy size sequences
$p^A=(p^A_i)$ and $p^B=(p^B_i)$ for the two sides, padded by zeros,
such that the greedy size sequence for $G$ is
\[
   p_i=p^A_i+p^B_i\qquad(i\geq 1).
\]
Both component sequences are nonincreasing, so $(p_i)$ is also
nonincreasing. Order the parts $Y_1,Y_2,\dots$ of $\mathcal{Y}$ by
nonincreasing size, and put $a_i=|Y_i\cap A|$ and
$b_i=|Y_i\cap B|$. The nonempty intersections form clique partitions
of $G[A]$ and $G[B]$. If $a^\downarrow$ and $b^\downarrow$ denote
the two intersection-size sequences in nonincreasing order, induction
gives, for every $j$,
  \begin{align*}
\sum_{i=1}^j |Y_i|
 &=\sum_{i=1}^j(a_i+b_i)\\
 &\leq \sum_{i=1}^j a^\downarrow_i+
       \sum_{i=1}^j b^\downarrow_i\\
 &\leq \sum_{i=1}^j(p^A_i+p^B_i)
 =\sum_{i=1}^j p_i.
  \end{align*}
Thus, $\mathcal{X}$ majorizes $\mathcal{Y}$ in both cases.
\end{proof}

Combining \Cref{lem:partialorder,thm:simpercograph} proves
\Cref{thm:cograph}.

\begin{rem} \label{rem:convex}
Since the majorization inequality holds for every convex function, \Cref{thm:simpercograph} gives more than \Cref{thm:cograph}.  Let $f$ be a function on the nonnegative integers with $f(0)=0$ and with $f(x+1)-f(x)$ nondecreasing in $x$. Extending $f$ piecewise linearly to $[0,\infty)$ gives a convex function, so Karamata's inequality applies to integer sequences.  Then, for every cograph $G$, every clique partition $\mathcal{X}$ output by \greedy{} maximizes $\sum_{X\in\mathcal{X}}f(|X|)$ over all clique partitions of $G$.  Taking $f(x)=\binom{x}{2}$ recovers \Cref{thm:cograph}. Taking $f(x)=-1$ for all $x\geq 1$ shows that \greedy{} also uses the minimum possible number of cliques. That is, it also computes the \emph{clique cover number} of a cograph.
\end{rem}

\subsection{A clique-partition linear program}

The preceding majorization argument actually shows that the natural linear
programming relaxation of \cp{} is exact on cographs. Recall that $\mathcal{C}(G)$ is the family of all
nonempty cliques of $G$, and that \eqref{lp:clique-partition} asks to maximize $\sum_{K\in\mathcal{C}(G)}\binom{|K|}{2}x_K$ subject to $x\geq 0$ and $\sum_{K \ni v} x_K=1$ for every $v\in V(G)$.
If the vertex constraints are instead written as inequalities, their
slack can be assigned to singleton cliques without changing the
objective.  The incidence vector of every clique partition is a feasible solution of \eqref{lp:clique-partition}, so the optimal value of \eqref{lp:clique-partition} is always at least the maximum number of edges in a clique partition of $G$.  We say that \eqref{lp:clique-partition} is \defn{exact} for $G$ if equality holds.

The feasible region of \eqref{lp:clique-partition} is not integral,
even for cographs. Indeed, if $G=K_3$ with vertex set $\{1,2,3\}$,
then
\[
x_{\{1,2\}}=x_{\{1,3\}}=x_{\{2,3\}}=\frac12
\]
and all other variables zero give a fractional feasible solution. The
three positive columns form the matrix
\[
\begin{pmatrix}
1&1&0\\
1&0&1\\
0&1&1
\end{pmatrix},
\]
whose determinant is $-2$. Thus, this solution is a basic feasible
solution. However, we will show that \eqref{lp:clique-partition} always has an optimal solution that is $\{0,1\}$-valued when $G$ is a cograph.

To prove this statement, we extend majorization to fractional clique
partitions. A \defn{weighted size profile} is a finite family
\[
\mathcal{Q}=\{(q_\alpha,\lambda_\alpha):\alpha\in I\},
\qquad q_\alpha\geq 0,\quad \lambda_\alpha\geq 0.
\]
Let $q^\downarrow:[0,\infty)\to\mathbb{R}_{\geq0}$ be the decreasing
step function in which the value $q_\alpha$ occupies an interval of
length $\lambda_\alpha$, padded by zeros, and define
\[
L_{\mathcal{Q}}(t):=\int_0^t q^\downarrow(s)\,ds.
\]
Equivalently, $L_{\mathcal{Q}}(t)$ is the maximum total size obtainable
by taking total weight $t$ from $\mathcal{Q}$, allowing a fractional
amount of an entry.

Every feasible solution $x$ of \eqref{lp:clique-partition} has the
weighted size profile
\[
\mathcal{Q}_x
   :=\{(|K|,x_K):K\in\mathcal{C}(G)\}.
\]
The covering constraints imply
\begin{equation} \label{eq:fractional-profile-sum}
\sum_{K\in\mathcal{C}(G)}|K|x_K
=\sum_{v\in V(G)}
  \sum_{\substack{K\in\mathcal{C}(G)\\v\in K}}x_K
=|V(G)|.
\end{equation}

\begin{lem} \label{lem:fractional-majorization}
Let $G$ be a cograph, let $\mathcal{P}$ be a clique partition output
by \greedy{}, and let $p_1\geq p_2\geq\cdots$ be its clique sizes,
padded by zeros. For every feasible solution $x$ of
\eqref{lp:clique-partition} and every $t\geq0$,
\[
L_{\mathcal{Q}_x}(t)
\leq \int_0^t p^\downarrow(s)\,ds,
\]
where $p^\downarrow$ is the step function that takes the value $p_i$
on the interval $[i-1,i)$.
\end{lem}

\begin{proof}
We proceed by induction on $|V(G)|$. The claim is immediate when
$|V(G)|\leq1$. Choose a nontrivial cograph decomposition $(A,B)$ as
in \Cref{lem:cographpartition}.

First suppose that $G=G[A]\sqcup G[B]$. Every clique is contained in
one side, so $\mathcal{Q}_x$ is the union of feasible fractional
profiles $\mathcal{Q}_A$ and $\mathcal{Q}_B$ for the two sides. The
restriction of the greedy run to each side is a greedy run; denote its
profiles by $p^A$ and $p^B$. By induction, for every $t\geq0$,
\[
\begin{aligned}
L_{\mathcal{Q}_x}(t)
 &=\max_{\substack{r,s\geq0\\r+s=t}}
     \bigl(L_{\mathcal{Q}_A}(r)+L_{\mathcal{Q}_B}(s)\bigr)\\
 &\leq\max_{\substack{r,s\geq0\\r+s=t}}
     \bigl(L_{p^A}(r)+L_{p^B}(s)\bigr)
 =L_p(t),
\end{aligned}
\]
where $p$ is the nonincreasing merge of $p^A$ and $p^B$.

Now suppose that $G=G[A]\vee G[B]$. For every clique $K$ of $G$, put
$K_A=K\cap A$ and $K_B=K\cap B$. Using weight $x_K$ on the entries
$|K_A|$ and $|K_B|$ gives feasible fractional profiles
$\mathcal{Q}_A$ and $\mathcal{Q}_B$ on the two sides; zero entries may
simply be retained. For any choice of total weight $t$, the contribution
of the entries $|K_A|+|K_B|$ is at most the sum of the $t$ largest
entries from the two component profiles. Hence,
\[
L_{\mathcal{Q}_x}(t)
\leq L_{\mathcal{Q}_A}(t)+L_{\mathcal{Q}_B}(t).
\]
There are greedy profiles $p^A$ and $p^B$, padded by zeros, such that
the greedy profile of the join satisfies
\[
p_i=p^A_i+p^B_i\qquad(i\geq1).
\]
Since both component profiles are nonincreasing, induction gives
\[
L_{\mathcal{Q}_x}(t)
\leq L_{p^A}(t)+L_{p^B}(t)
=L_p(t).
\]
This completes the induction.
\end{proof}

We are now ready to prove~\Cref{thm:cograph-lp}.

\begin{proof}[Proof of~\Cref{thm:cograph-lp}]
Let $x$ be any feasible solution, and let $\mathcal{P}$ be a clique
partition output by \greedy{}, with size sequence
$p_1\geq p_2\geq\cdots$. By
\Cref{eq:fractional-profile-sum,lem:fractional-majorization}, the
profile $p$ majorizes the weighted profile $\mathcal{Q}_x$, and both
profiles have total size $|V(G)|$. The weighted form of Karamata's
inequality (see~\cite{MOA11}), applied to the convex function
$f(z)=\binom{z}{2}$, therefore gives
\[
\sum_{K\in\mathcal{C}(G)}x_K\binom{|K|}{2}
\leq \sum_i\binom{p_i}{2}
=|E(\mathcal{P})|.
\]
The incidence vector of $\mathcal{P}$ is a feasible integral solution
of \eqref{lp:clique-partition}. It is therefore at least as good as
every fractional solution, and hence is optimal.
\end{proof}

We conclude this section by showing that \Cref{thm:cograph-lp} does not extend to permutation graphs.  We used Claude Fable 5.1 to find this example, and to verify by exhaustive computer search that it is the smallest such example.

\begin{prop} \label{prop:lp-perm}
Let $\pi=[5,9,8,7,3,6,2,1,4]$ and let $G=G_\pi$ be the corresponding permutation graph, depicted in \cref{fig:lp-perm}.  Then the optimal value of \eqref{lp:clique-partition} for $G$ is $33/2$, whereas every clique partition of $G$ has at most $16$ edges.
\end{prop}

\begin{figure}[h!]
\centering
\begin{tikzpicture}[scale=.9]
    \tikzstyle{vertex}=[circle,fill=white, draw=black, inner sep=0.cm, minimum size=4.5mm]
	\begin{scope}
		\node[vertex] (1) at (0,0){{\tiny $1$}};
		\node[vertex] (2) at (1.2,0.9){{\tiny $2$}};
		\node[vertex] (7) at (2.4,0){{\tiny $7$}};
		\node[vertex] (8) at (1.9,-1.4){{\tiny $8$}};
		\node[vertex] (9) at (0.5,-1.4){{\tiny $9$}};
		\node[vertex] (3) at (-1.6,0.9){{\tiny $3$}};
		\node[vertex] (6) at (4.0,0.9){{\tiny $6$}};
		\node[vertex] (5) at (-1.6,2.7){{\tiny $5$}};
		\node[vertex] (4) at (4.0,2.7){{\tiny $4$}};
		\draw (1)--(2); \draw (1)--(7); \draw (1)--(8); \draw (1)--(9);
		\draw (2)--(7); \draw (2)--(8); \draw (2)--(9);
		\draw (7)--(8); \draw (7)--(9); \draw (8)--(9);
		\draw (3)--(1); \draw (3)--(2); \draw (3)--(7); \draw (3)--(8); \draw (3)--(9);
		\draw (6)--(1); \draw (6)--(2); \draw (6)--(7); \draw (6)--(8); \draw (6)--(9);
		\draw (5)--(3); \draw (5)--(1); \draw (5)--(2);
		\draw (4)--(6); \draw (4)--(7); \draw (4)--(8); \draw (4)--(9);
		\draw (5)--(4);
	\end{scope}
\end{tikzpicture}
\caption{The permutation graph $G$ of $[5,9,8,7,3,6,2,1,4]$, for which \eqref{lp:clique-partition} is not exact.}
\label{fig:lp-perm}
\end{figure}

\begin{proof}
Recall that the cliques of $G=G_\pi$ are the decreasing subsequences of $\pi$.  One checks directly that the maximal cliques of $G$ are
\[
   \{1,2,3,7,8,9\},\quad \{1,2,6,7,8,9\},\quad \{1,2,3,5\},\quad \{4,6,7,8,9\},\quad \{4,5\},
\]
and that $G$ has $28$ edges.  Equivalently, $G$ is obtained from the complete graph on $\{1,2,3,6,7,8,9\}$ by deleting the edge $\{3,6\}$ and adding two vertices $4$ and $5$, where $5$ is adjacent to $1,2,3,4$ and $4$ is adjacent to $5,6,7,8,9$.  In particular, $\omega(G)=6$, and the only $6$-cliques are the first two sets above.

Assign $x_K=1/2$ to each of the five cliques
\[
  \{1,2,3,7,8,9\},\quad \{1,2,6,7,8,9\},\quad \{3,5\},\quad \{4,5\},\quad \{4,6\},
\]
and $x_K=0$ to all other cliques.  Every vertex is covered exactly once, so $x$ is feasible for \eqref{lp:clique-partition}, and its objective value is $\frac12(15+15+1+1+1)=33/2$.  On the other hand, the dual of \eqref{lp:clique-partition} asks to minimize $\sum_{v\in V(G)}y_v$ subject to $\sum_{v\in K}y_v\geq\binom{|K|}{2}$ for every $K\in\mathcal{C}(G)$. The vector $y$ with
\[
   y_1=3,\; y_2=2,\; y_3=y_4=y_5=y_6=\tfrac12,\; y_7=\tfrac92,\; y_8=2,\; y_9=3
\]
is dual feasible.  Indeed, every clique of $G$ is contained in one of the five maximal cliques $Q$ listed above, so it suffices to check that for each such $Q$ and each $k\leq|Q|$, the sum of the $k$ smallest values $y_v$ ($v\in Q$) is at least $\binom{k}{2}$.  For $Q=\{1,2,3,7,8,9\}$ and $Q=\{1,2,6,7,8,9\}$, the values $y_v$ in nondecreasing order are $\frac12,2,2,3,3,\frac92$, with partial sums $\frac12,\frac52,\frac92,\frac{15}{2},\frac{21}{2},15$; for $Q=\{4,6,7,8,9\}$ they are $\frac12,\frac12,2,3,\frac92$, with partial sums $\frac12,1,3,6,\frac{21}{2}$; for $Q=\{1,2,3,5\}$ they are $\frac12,\frac12,2,3$, with partial sums $\frac12,1,3,6$; and for $Q=\{4,5\}$ the partial sums are $\frac12,1$.  In each case the $k$-th partial sum is at least $\binom{k}{2}$.  Since $\sum_v y_v=33/2$, linear programming duality shows that the optimal value of \eqref{lp:clique-partition} is exactly $33/2$.

Finally, let $\mathcal{X}$ be a clique partition of $G$ with part sizes $s_1\geq s_2\geq\cdots$. Since $\sum_j s_j=9$ and $s_1\leq 6$, a direct enumeration of the partitions of $9$ into parts of size at most $6$ shows that $\sum_j\binom{s_j}{2}\geq 17$ forces $(s_1,s_2)=(6,3)$, that is, $\mathcal{X}$ consists of a $6$-clique and a disjoint triangle.  However, $V(G)\setminus\{1,2,3,7,8,9\}=\{4,5,6\}$ and $V(G)\setminus\{1,2,6,7,8,9\}=\{3,4,5\}$ are not cliques.  Hence $|E(\mathcal{X})|\leq 16$, and the clique partition $\{\{1,2,3,7,8,9\},\{4,6\},\{5\}\}$ shows that this bound is attained.
\end{proof}

By \Cref{thm:NPhardness} below, the failure of exactness on permutation graphs is not an accident. The dual of \eqref{lp:clique-partition} can be solved in polynomial time on permutation graphs as follows.  Since the cliques of $G_\pi$ of size $q$ are the decreasing subsequences of $\pi$ of length $q$, a minimum $y$-weight clique of each size $q\in[n]$ can be found by dynamic programming, which yields a polynomial-time separation oracle. Thus, \eqref{lp:clique-partition} can be solved in polynomial time via the ellipsoid method~\cite{GLS93}.  Consequently, if \eqref{lp:clique-partition} were exact for all permutation graphs, then the optimal value of \cp{} on permutation graphs could be computed in polynomial time, which by \Cref{thm:NPhardness} is impossible unless $\mathsf{P}=\mathsf{NP}$.  

\section{Cluster deletion in permutation graphs is NP-complete} \label{sec:NPhard}

In this section, we prove \Cref{thm:NPhardness}, which resolves an open problem of Konstantinidis and Papadopoulos~\cite{konstantinidis2021cluster}.  As explained in the introduction, the decision versions of \cp{} and \cd{} are equivalent and belong to $\mathsf{NP}$, so it suffices to show that \cp{} is NP-hard on permutation graphs.

We give a polynomial reduction from the \occp{} (OCCP) problem.

\medskip

\noindent\fbox{\parbox{\dimexpr\linewidth-2\fboxsep-2\fboxrule-4pt\relax}{%
$\occp$:
\\
{\em Input:} A graph $H$ with $n$ vertices and nonnegative integers $k_1 \leq k_2 \leq \dots \leq k_n$.\\
{\em Output:} A partition $\mathcal{S}=(S_1,\dots,S_n)$ of $V(H)$ into stable sets of $H$ (some of which may be empty) such that $\cost(\mathcal{S}) := \sum_{i =1}^n k_i |S_i|$ is minimum.}
}

\medskip

\noindent In other words, color $i$ has cost $k_i$, and OCCP asks for a proper coloring of $H$ of minimum total cost; since $H$ has only $n$ vertices, allowing $n$ colors is not a restriction.  We call a partition of $V(H)$ into $n$ stable sets a \defn{coloring} of $H$.  In the decision version of OCCP, we are additionally given an integer $c$, and the question is whether $H$ has a coloring of cost at most $c$.  Jansen~\cite{JANSEN200054} proved that OCCP is NP-complete on permutation graphs.  Moreover, in the instances constructed in his hardness proof, the costs $k_i$ are bounded by a polynomial in $n$ (see~\cite[Section 4]{JANSEN200054}).  We only use this restricted form of Jansen's result.

\begin{thm}[Jansen~\cite{JANSEN200054}] \label{thm:jansen}
    There is a constant $d$ such that the decision version of OCCP is NP-hard when restricted to instances $(H, (k_1, \dots, k_n), c)$ in which $H$ is a permutation graph and $k_n \leq n^d$.
\end{thm}

\subsection{Scores}

It is convenient to measure clique partitions by a slightly different quantity than the number of edges.  For a clique partition $\mathcal{K}=(K_1, \dots, K_m)$ of a graph $G$, we define the \defn{score} of $\mathcal{K}$ to be
\[
    \score(\mathcal{K}) := \sum_{j=1}^m |K_j|^2.
\]
Since $\sum_{j=1}^m |K_j| = |V(G)|$, we have
\begin{equation} \label{eq:score}
    |E(\mathcal{K})| = \sum_{j=1}^m \binom{|K_j|}{2} = \frac{\score(\mathcal{K}) - |V(G)|}{2}.
\end{equation}
Thus, maximizing $|E(\mathcal{K})|$ is the same as maximizing $\score(\mathcal{K})$, and for every integer $t$, $G$ has a clique partition with at least $t$ edges if and only if $G$ has a clique partition of score at least $2t+|V(G)|$.

\subsection{The construction}

Fix an instance $(H, (k_1, \dots, k_n), c)$ of OCCP, where $H$ is a graph with vertex set $V(H)$ of size $n$, and $k_1 \leq \dots \leq k_n$ are nonnegative integers.  Put
\[
    S := \sum_{i=1}^n k_i, \qquad M := n^2+1, \qquad a_i := 1 + M(S-k_i) \quad (i \in [n]).
\]
Note that $a_1 \geq a_2 \geq \dots \geq a_n \geq 1$.  Let $A_1, \dots, A_n$ be pairwise disjoint sets of vertices, disjoint from $V(H)$, with $|A_i|=a_i$ for all $i \in [n]$, and let
\[
    G := \overline{H} \vee \bigl(K[A_1] \sqcup K[A_2] \sqcup \dots \sqcup K[A_n]\bigr),
\]
where $K[A_i]$ denotes the complete graph with vertex set $A_i$.  Explicitly, $V(G)=V(H) \cup A_1 \cup \dots \cup A_n$; two vertices of $H$ are adjacent in $G$ if and only if they are nonadjacent in $H$; every vertex of $H$ is adjacent to every vertex of $A_1 \cup \dots \cup A_n$; each $A_i$ is a clique of $G$; and there are no edges between $A_i$ and $A_j$ for $i \neq j$.  Finally, put
\begin{equation} \label{eq:C}
    C := \sum_{i=1}^n a_i^2 + 2n + 2MSn
    \qquad\text{and}\qquad
    T := C - 2Mc.
\end{equation}

We record two simple observations. First, since there are no edges between distinct sets $A_i$ and $A_j$, and since the cliques of $\overline{H}$ are the stable sets of $H$, we have the following description of the cliques of $G$.

\begin{fact} \label{fact:cliques-of-G}
    The cliques of $G$ are exactly the sets of the form $B \cup Q$, where $B \subseteq A_i$ for some $i\in[n]$ and $Q$ is a stable set of $H$.  In particular, no clique of $G$ meets two of the sets $A_1, \dots, A_n$.
\end{fact}

Second, $|V(G)| = n + \sum_{i=1}^n a_i \leq n + n(1+MS) = O(n^3 S)$.  Hence, if $k_n \leq n^d$, then $S \leq n^{d+1}$ and $|V(G)|=O(n^{d+4})$, so $G$ can be constructed in time polynomial in $n$.

For a coloring $\mathcal{S}=(S_1, \dots, S_n)$ of $H$, we define
\[
    \mathcal{K}_{\mathcal{S}} := (A_1 \cup S_1, \dots, A_n \cup S_n).
\]

\begin{lem} \label{lem:score-of-coloring}
    For every coloring $\mathcal{S}=(S_1, \dots, S_n)$ of $H$, $\mathcal{K}_{\mathcal{S}}$ is a clique partition of $G$, and
    \[
        \score(\mathcal{K}_{\mathcal{S}}) = C + \sum_{i=1}^n |S_i|^2 - 2M \cdot \cost(\mathcal{S}).
    \]
\end{lem}

\begin{proof}
    Each $A_i \cup S_i$ is a clique of $G$ by~\Cref{fact:cliques-of-G}, and these sets partition $V(G)$.  Using $\sum_{i=1}^n |S_i| = n$, we compute
    \begin{align*}
        \score(\mathcal{K}_{\mathcal{S}})
        &= \sum_{i=1}^n (a_i + |S_i|)^2
         = \sum_{i=1}^n a_i^2 + \sum_{i=1}^n |S_i|^2 + 2\sum_{i=1}^n a_i |S_i| \\
        &= \sum_{i=1}^n a_i^2 + \sum_{i=1}^n |S_i|^2 + 2\sum_{i=1}^n \bigl(1 + M(S-k_i)\bigr)|S_i| \\
        &= \sum_{i=1}^n a_i^2 + \sum_{i=1}^n |S_i|^2 + 2n + 2MSn - 2M\sum_{i=1}^n k_i|S_i| \\
        &= C + \sum_{i=1}^n |S_i|^2 - 2M \cdot \cost(\mathcal{S}). \qedhere
    \end{align*}
\end{proof}

The next lemma shows that, as far as maximum scores are concerned, we may restrict our attention to clique partitions of the form $\mathcal{K}_{\mathcal{S}}$.

\begin{lem} \label{lem:structure}
    Let $\mathcal{K}$ be a clique partition of $G$ of maximum score.  Then $\mathcal{K}=\mathcal{K}_{\mathcal{S}}$ for some coloring $\mathcal{S}$ of $H$.  Consequently, for every clique partition $\mathcal{K}'$ of $G$ there is a coloring $\mathcal{S}$ of $H$ with $\score(\mathcal{K}_{\mathcal{S}}) \geq \score(\mathcal{K}')$.
\end{lem}

\begin{proof}
    We first show that each $A_i$ is contained in a single part of $\mathcal{K}$.  Suppose not, and let $K$ and $K'$ be distinct parts of $\mathcal{K}$ that both meet $A_i$.  By~\Cref{fact:cliques-of-G}, $K = B \cup Q$ and $K'=B' \cup Q'$, where $B$ and $B'$ are disjoint nonempty subsets of $A_i$, and $Q, Q' \subseteq V(H)$ are stable sets of $H$.  By symmetry, we may assume that $|Q| \geq |Q'|$.  Let $\mathcal{K}^*$ be obtained from $\mathcal{K}$ by replacing $K$ and $K'$ by $K \cup B' = B \cup B' \cup Q$ and $K' \setminus B' = Q'$ (and discarding $Q'$ if it is empty).  By~\Cref{fact:cliques-of-G}, $K \cup B'$ is a clique of $G$, so $\mathcal{K}^*$ is a clique partition of $G$, and
    \begin{align*}
        \score(\mathcal{K}^*) - \score(\mathcal{K})
        &= \bigl(|B|+|B'|+|Q|\bigr)^2 + |Q'|^2 \\
        &\quad - \bigl(|B|+|Q|\bigr)^2 - \bigl(|B'|+|Q'|\bigr)^2 \\
        &= 2|B'|\bigl(|B| + |Q| - |Q'|\bigr) > 0,
    \end{align*}
    since $|B'| \geq 1$, $|B| \geq 1$ and $|Q| \geq |Q'|$.  This contradicts the maximality of $\score(\mathcal{K})$.

    Hence, for each $i \in [n]$ there is a unique part $K_i$ of $\mathcal{K}$ containing $A_i$, and by~\Cref{fact:cliques-of-G} we can write $K_i = A_i \cup S_i$ with $S_i \subseteq V(H)$ a stable set of $H$.  The parts $K_1, \dots, K_n$ are distinct, since by~\Cref{fact:cliques-of-G} no clique of $G$ meets two of the sets $A_i$, and every other part of $\mathcal{K}$ is a nonempty subset of $V(H)$.  Suppose that there is such a part $K_0 \subseteq V(H)$.  Then $\sum_{i=1}^n |S_i| \leq n - |K_0| < n$, so $S_i = \emptyset$ for some $i \in [n]$, that is, $K_i = A_i$.  By~\Cref{fact:cliques-of-G}, $A_i \cup K_0$ is a clique of $G$, and replacing the two parts $K_i$ and $K_0$ by the single part $A_i \cup K_0$ increases the score by $(a_i+|K_0|)^2 - a_i^2 - |K_0|^2 = 2a_i|K_0| > 0$, again contradicting the maximality of $\score(\mathcal{K})$.

    Therefore, $\mathcal{K} = (A_1 \cup S_1, \dots, A_n \cup S_n) = \mathcal{K}_{\mathcal{S}}$, where $\mathcal{S}=(S_1, \dots, S_n)$ is a coloring of $H$.  The last assertion follows since a clique partition of maximum score exists.
\end{proof}

We are now ready to establish the correctness of the reduction.

\begin{thm} \label{thm:reduction}
    $H$ has a coloring of cost at most $c$ if and only if $G$ has a clique partition of score at least $T$.
\end{thm}

\begin{proof}
    Suppose first that $\mathcal{S}=(S_1, \dots, S_n)$ is a coloring of $H$ with $\cost(\mathcal{S}) \leq c$.  By \Cref{lem:score-of-coloring},
    \[
        \score(\mathcal{K}_{\mathcal{S}}) = C + \sum_{i=1}^n |S_i|^2 - 2M\cdot \cost(\mathcal{S}) \geq C - 2Mc = T.
    \]
    Conversely, suppose that $G$ has a clique partition of score at least $T$.  By \Cref{lem:structure}, there is a coloring $\mathcal{S}=(S_1, \dots, S_n)$ of $H$ with $\score(\mathcal{K}_{\mathcal{S}}) \geq T$.  By \Cref{lem:score-of-coloring},
    \[
        2M\bigl(\cost(\mathcal{S}) - c\bigr) \leq \sum_{i=1}^n |S_i|^2 \leq \Bigl(\sum_{i=1}^n |S_i|\Bigr)^2 = n^2 < 2M.
    \]
    Since $\cost(\mathcal{S})$ and $c$ are integers, it follows that $\cost(\mathcal{S}) \leq c$.
\end{proof}

The same computation shows that the reduction maps optimal solutions to optimal solutions. Indeed, let $\mathcal{K}$ be a clique partition of $G$ of maximum score. By \Cref{lem:structure}, $\mathcal{K}=\mathcal{K}_{\mathcal{S}}$ for some coloring $\mathcal{S}$ of $H$, and if some coloring $\mathcal{S}'$ of $H$ had $\cost(\mathcal{S}') < \cost(\mathcal{S})$, then \Cref{lem:score-of-coloring} would give $\score(\mathcal{K}_{\mathcal{S}'}) - \score(\mathcal{K}_{\mathcal{S}}) \geq 2M - n^2 > 0$, a contradiction.  Thus, $\mathcal{S}$ is a minimum cost coloring of $H$ (in fact, one maximizing $\sum_{i=1}^n |S_i|^2$ among all minimum cost colorings).

\subsection{Preserving permutation graphs}

It remains to show that the reduction preserves the property of being a permutation graph.  This also follows from the facts that the class of permutation graphs is closed under complementation, disjoint unions and joins (see~\cite{golumbic2004}), but we give an explicit permutation.

\begin{lem} \label{lem:G-permutation}
    If $H$ is a permutation graph, then so is $G$.  Moreover, given a permutation $\pi$ with $H \cong G_\pi$, a permutation $\sigma$ with $G \cong G_\sigma$ can be computed in time polynomial in $|V(G)|$.
\end{lem}

\begin{proof}
    We may assume that $V(H)=[n]$ and $H = G_\pi$ for a permutation $\pi=[\pi_1, \dots, \pi_n]$ of $[n]$.  Put $N := \sum_{i=1}^n a_i$ and $n_i := a_1 + \dots + a_i$ for $i \in \{0, 1, \dots, n\}$, so that $n_0 = 0$ and $n_n=N$.  Let $\sigma$ be the following permutation of $[N+n]$, written as a word:
    \[
    \begin{aligned}
        \sigma := [\,&\pi_n + N,\ \pi_{n-1}+N,\ \dots,\ \pi_1+N,\\
        &n_1,\ n_1-1,\ \dots,\ n_0+1,\\
        &n_2,\ n_2-1,\ \dots,\ n_1+1,\\
        &\ \ \vdots\\
        &n_n,\ n_n-1,\ \dots,\ n_{n-1}+1\,].
    \end{aligned}
    \]
    That is, $\sigma$ begins with the reverse of $\pi$ with all values shifted by $N$, followed by $n$ decreasing runs, the $i$-th of which consists of the values in $(n_{i-1}, n_i]$.  We identify the vertex $v \in V(H)$ with the value $v+N$, and the set $A_i$ with the set of values in $(n_{i-1}, n_i]$, and claim that under this identification $G_\sigma = G$.  Recall that two values are adjacent in $G_\sigma$ if and only if the smaller value appears later in the word $\sigma$.
    \begin{itemize}
        \item Let $u < v$ be vertices of $H$.  Then $u+N$ appears later than $v+N$ in $\sigma$ if and only if $u$ appears earlier than $v$ in $\pi$, that is, if and only if $u$ and $v$ are nonadjacent in $H=G_\pi$.  Hence $G_\sigma[V(H)] = \overline{H}$.
        \item Let $u \in V(H)$ and $x \in A_i$.  Then $x < u+N$ and $x$ appears later than $u+N$ in $\sigma$, so $x$ and $u$ are adjacent in $G_\sigma$.
        \item Let $x < y$ be values in $A_1 \cup \dots \cup A_n$.  If $x, y \in A_i$ for some $i$, then $x$ appears later than $y$ (the $i$-th run is decreasing), so $x$ and $y$ are adjacent.  If $x \in A_i$ and $y \in A_j$ with $i \neq j$, then $i<j$, and $x$ appears earlier than $y$, so $x$ and $y$ are nonadjacent.
    \end{itemize}
    Therefore, $G_\sigma = G$.  Clearly, $\sigma$ can be written down in time polynomial in $N+n=|V(G)|$.
\end{proof}

We can now complete the proof of \Cref{thm:NPhardness}.

\begin{proof}[Proof of \Cref{thm:NPhardness}]
    As noted above, the decision versions of \cp{} and \cd{} are in $\mathsf{NP}$ and are polynomially equivalent, so it suffices to show that \cp{} is NP-hard on permutation graphs.  Let $(H, (k_1, \dots, k_n), c)$ be an instance of OCCP with $H$ a permutation graph and $k_n \leq n^d$, where $d$ is the constant from \Cref{thm:jansen}.  Construct the graph $G$ and the integer $T$ as in~\eqref{eq:C}, and put $t := \lceil (T - |V(G)|)/2 \rceil$.  A permutation $\pi$ with $H \cong G_\pi$ can be found in linear time~\cite{MS99}, so since $|V(G)|=O(n^{d+4})$, \Cref{lem:G-permutation} shows that the permutation graph $G$ (together with a permutation $\sigma$ representing it) can be computed in polynomial time.  By \Cref{thm:reduction} and~\eqref{eq:score}, $H$ has a coloring of cost at most $c$ if and only if $G$ has a clique partition with at least $t$ edges.  By \Cref{thm:jansen}, \cp{} is NP-hard on permutation graphs.
\end{proof}

Since the permutation graphs are precisely the graphs that are simultaneously comparability graphs and cocomparability graphs (see~\cite{golumbic2004}), \Cref{thm:NPhardness} immediately yields hardness for these two well-studied superclasses.  Note that both classes contain natural subclasses on which \cd{} is polynomial-time solvable. Bipartite graphs are comparability graphs (and on triangle-free graphs \cd{} is just a maximum matching problem), while interval graphs are cocomparability graphs, and \cd{} is polynomial-time solvable on interval graphs~\cite{BONOMO2015,konstantinidis2021cluster}.

\begin{cor} \label{cor:comparability}
    \cd{} and \cp{} are NP-complete on comparability graphs and on cocomparability graphs.
\end{cor}

\section{An improved approximation algorithm}
 In this section we present an improved approximation algorithm for \cp.  The description of the algorithm is simple. While the current graph has clique number at least $3$, we compute a maximum clique $X$, put $X$ into the clique partition, and delete $X$. Once the remaining graph has clique number at most $2$, an optimal clique partition consists of a maximum matching together with singleton cliques. We call the algorithm $\edmonds$.

\begin{algorithm}[h!]
\caption{\edmonds}
\begin{tabularx}{\columnwidth}{@{}rX@{}}
\textbf{Input:} & A graph $G$.\\
\textbf{Output:} & A clique partition $\mathcal X$ of $G$.\\[2pt]
1. & Set $\mathcal X=\emptyset$.\\
2. & While $\omega(G)\geq 3$:\\
   & \quad choose a maximum clique $X$ of $G$;\\
   & \quad set $\mathcal X\gets\mathcal X\cup\{X\}$ and $G\gets G-X$.\\
3. & Compute a maximum matching $\mathcal M$ of $G$.\\
4. & Add the edges of $\mathcal M$ as two-vertex cliques and all
uncovered vertices as singleton cliques.\\
5. & Return $\mathcal X$.
\end{tabularx}
\end{algorithm}





Let $G$ be a graph, $\mathcal{Y}$ be a clique partition of $G$, and $C \subseteq V(G)$.   We let $\mathcal{Y}-C$ be the collection of nonempty sets of the form $Y\setminus C$, where $Y \in \mathcal{Y}$.  Note that $\mathcal{Y}-C$ is a clique partition of $G-C$.  Recall that $\omega'(G)$ is the number of edges in a maximum clique of $G$.

We now show that \edmonds{} is a $\frac{2\omega'(G)}{\omega'(G)+1}$-approximation algorithm for every graph $G$.
We first require the following lemma.

\begin{lem} \label{lem:ordering}
    Let $G$ be a graph, $\mathcal{Y}$ be a clique partition of $G$,
    $\lambda(k):=\min\{|E(\mathcal{Y}-X)|: X\subseteq V(G),\, |X|=k\}$, and $C$ be a subset of $V(G)$ of size $k$.  If $|E(\mathcal{Y}-C)|=\lambda(k)$, then there exists an ordering $c_1, \dots, c_k$ of $C$ such that $c_i$ lies in a largest part of $\mathcal{Y}-\{c_1, \dots, c_{i-1}\}$ for all $i \in [k]$.  
\end{lem}

\begin{proof}
We proceed by induction on $k$. The assertion is immediate for $k=1$.
Suppose $k\geq 2$, and let $C$ attain $\lambda(k)$. We first show
that $C$ contains a vertex in a largest part of $\mathcal{Y}$. If it
does not, choose $x\in C$ and choose $y$ in a largest part of
$\mathcal{Y}$. Put $D=C\setminus\{x\}$ and
$C'=D\cup\{y\}$. In the partition $\mathcal{Y}-D$, the part
containing $y$ has its original maximum size, say $q$, whereas the
part containing $x$ has size at most $q-1$. Deleting a vertex from a
part of size $s$ removes exactly $s-1$ retained edges. Therefore,
\[
   |E(\mathcal{Y}-C')|<|E(\mathcal{Y}-C)|,
\]
contrary to the choice of $C$.

Choose such a vertex $c_1\in C$, put
$C_1=C\setminus\{c_1\}$, and let
$\mathcal{Y}_1=\mathcal{Y}-\{c_1\}$. The set $C_1$ minimizes
$|E(\mathcal{Y}_1-D)|$ among all $(k-1)$-subsets
$D\subseteq V(G)\setminus\{c_1\}$; otherwise, adjoining $c_1$ to a
better set $D$ would contradict the minimality of $C$. The induction
hypothesis applied to $(\mathcal{Y}_1,C_1)$ gives an ordering
$c_2,\ldots,c_k$ with the required property. Prepending $c_1$
completes the proof.
\end{proof}

\begin{thm} \label{edmonds1}
    For every graph $G$, \edmonds{} returns a clique partition $\mathcal{X}$ such that \[\opt \leq \frac{2\omega'(G)}{\omega'(G)+1}|E(\mathcal{X})|,\] where $\opt$ is the number of edges in an optimal clique partition of $G$. 
\end{thm}

\begin{proof}
We proceed by induction on $|V(G)|$. Put
\[
   b_j:=\binom{j}{2}
   \quad\text{and}\quad
   \gamma_j:=\frac{b_j+1}{2b_j}\quad(j\geq 2).
\]
Let $\ell=\omega(G)$. If $\ell\leq 2$, then \edmonds{} is optimal,
so assume $\ell\geq 3$. Let $\mathcal{X}$ be the partition returned
by \edmonds{}, let $\mathcal{Y}$ be an optimal clique partition, and
let $x_i$ and $y_i$ be the numbers of parts of size $i$ in
$\mathcal{X}$ and $\mathcal{Y}$, respectively.

All $x_\ell$ cliques of size $\ell$ chosen by the algorithm occur
before any smaller chosen clique. Let $A$ be their union, put
$a=|A|=\ell x_\ell$, and set
\[
   G'=G-A,\qquad
   \mathcal{X}'=\mathcal{X}-A,\qquad
   \mathcal{Y}'=\mathcal{Y}-A.
\]
Then $\omega(G')\leq\ell-1$, and $\mathcal{X}'$ is an output of
\edmonds{} on $G'$. The induction hypothesis, together with the fact
that $\gamma_j$ decreases with $j$, gives
\begin{equation} \label{eq:induction-remainder}
   |E(\mathcal{X}')|\geq \gamma_{\ell-1}|E(\mathcal{Y}')|,
\end{equation}
For $\ell=3$ this uses $\gamma_2=1$, since $G'$ has clique number at
most $2$ and the algorithm is optimal.

Choose an $a$-element set $C\subseteq V(G)$ minimizing
$|E(\mathcal{Y}-C)|$. Since $A$ is also an $a$-element set,
\[
   |E(\mathcal{Y}')|\geq |E(\mathcal{Y}-C)|.
\]
The set $A$ meets every part of $\mathcal{Y}$ of size $\ell$;
otherwise that part would be an $\ell$-clique in $G'$. In particular,
$a\geq y_\ell$. By \Cref{lem:ordering}, the vertices of $C$ can be
removed successively from a currently largest part of
$\mathcal{Y}$. The first $y_\ell$ removals therefore meet distinct
parts of size $\ell$, and no part of size $\ell$ remains in
$\mathcal{Y}-C$.

If $C=V(G)$, then
\begin{align*}
   |E(\mathcal{Y})|
   &=\sum_{i=1}^{\ell}b_i y_i\\
   &\leq \frac{\ell-1}{2}\sum_{i=1}^{\ell}i y_i\\
   &=\frac{\ell-1}{2}a
   =b_\ell x_\ell
   \leq |E(\mathcal{X})|,
\end{align*}
and we are done. We may therefore let $k$ be the largest size of a
part in the nonempty partition $\mathcal{Y}-C$. Then
$1\leq k\leq\ell-1$. The greedy deletion order supplied by
\Cref{lem:ordering} first reduces every original part of size greater
than $k$ to size $k$. This uses
\[
   s:=\sum_{i=k}^{\ell}(i-k)y_i
\]
vertices. The remaining $d:=a-s\geq 0$ deletions reduce $d$ of the
resulting parts of size $k$ to size $k-1$ (when $k=1$, they remove
those singleton parts). Consequently,
\begin{align}
|E(\mathcal{Y}-C)|
 &=b_k\left(\sum_{i=k}^{\ell}y_i-d\right)
   +b_{k-1}d \notag\\
 &\quad+\sum_{i=1}^{k-1}b_i y_i \notag\\
 &=a(1-k)
   +\sum_{i=k}^{\ell}
      \frac{(2i-k)(k-1)}{2}y_i \notag\\
 &\quad
   +\sum_{i=1}^{k-1}b_i y_i. \label{eq:minimum-remainder}
\end{align}
This makes explicit the part sizes of $\mathcal{Y}-C$ determined by
the deletion order supplied by \Cref{lem:ordering}.

By \eqref{eq:induction-remainder} and the choice of $C$,
\begin{equation} \label{eq:algorithm-lower-bound}
|E(\mathcal{X})|
\geq b_\ell x_\ell+\gamma_{\ell-1}|E(\mathcal{Y}-C)|.
\end{equation}
Our goal is $|E(\mathcal{X})|\geq\gamma_\ell|E(\mathcal{Y})|$.
Because $\gamma_{\ell-1}>\gamma_\ell$, the terms with indices
$i<k$ in \eqref{eq:minimum-remainder} already have a sufficiently
large coefficient. Thus, it is enough to prove
\begin{equation} \label{eq:reduced-target}
\begin{split}
b_\ell x_\ell+\gamma_{\ell-1}\biggl(
a(1-k)&+
\sum_{i=k}^{\ell}\frac{(2i-k)(k-1)}{2}y_i\biggr)\\
&\geq\gamma_\ell\sum_{i=k}^{\ell}b_i y_i.
\end{split}
\end{equation}
Since $a=\ell x_\ell$,
\[
\theta:=\frac{(\ell-k)b_{\ell-1}-k+1}{2b_{\ell-1}}.
\]
\begin{align*}
b_\ell x_\ell+\gamma_{\ell-1}a(1-k)
 &=\theta a\\
 &\geq \theta\sum_{i=k}^{\ell}(i-k)y_i.
\end{align*}
Here we used $a\geq s$. The coefficient $\theta$ is nonnegative: as a
function of $k\leq\ell-1$, its numerator is minimized at
$k=\ell-1$, where it equals
$b_{\ell-1}-\ell+2=(\ell-2)(\ell-3)/2$.

It remains to check, for $k\leq i\leq\ell$, that
\begin{equation*}
\begin{split}
f(i,k,\ell):={}&
\frac{(\ell-k)b_{\ell-1}-k+1}{2b_{\ell-1}}(i-k)\\
&+\gamma_{\ell-1}\frac{(2i-k)(k-1)}{2}
-\gamma_\ell b_i
\geq 0.
\end{split}
\tag{$*$}
\end{equation*}
Indeed, summing $(*)$ with weights $y_i$ gives
\eqref{eq:reduced-target}. Clearing the positive denominator
$4\ell(\ell-1)(\ell-2)$ shows that the sign of $f$ is the sign of
\begin{align*}
g(i,k,\ell):={}&
(-\ell^3+3\ell^2-4\ell+4)i^2\\
&+(2\ell^4-7\ell^3+7\ell^2-4)i\\
&+\ell k\bigl(-2\ell^3+\ell^2k+7\ell^2\\
&\hspace{4.4em}{}-3\ell k-7\ell+4k\bigr).
\end{align*}
For fixed $k,\ell$, this is a concave quadratic in $i$, because its
leading coefficient is
$-(\ell-2)(\ell^2-\ell+2)<0$. Hence it suffices to check the
endpoints $i=k$ and $i=\ell$. Direct simplification gives
\[
   g(k,k,\ell)=4k(k-1)\geq 0.
\]
Moreover, $g(\ell,k,\ell)=\ell\,h(k,\ell)$, where
\[
\begin{split}
h(k,\ell):={}&(\ell^2-3\ell+4)k^2-(2\ell^3-7\ell^2+7\ell)k\\
&+\ell^4-4\ell^3+3\ell^2+4\ell-4.
\end{split}
\]
As a polynomial in $k$, $h(k,\ell)$ has positive leading
coefficient $\ell^2-3\ell+4$ and roots
\[
   r_1=\ell-1,\qquad
   r_2=\frac{(\ell-2)^2(\ell+1)}{\ell^2-3\ell+4}.
\]
For $\ell\geq 6$, we have $r_2>r_1$, since
$r_2-r_1=\frac{\ell^2-7\ell+8}{\ell^2-3\ell+4}>0$; hence every
$1\leq k\leq\ell-1$ lies at or to the left of both roots. For
$\ell\in\{3,4,5\}$, we have
$\ell-2\leq r_2<r_1=\ell-1$, since
$r_2-(\ell-2)=\frac{2(\ell-2)(\ell-3)}{\ell^2-3\ell+4}\geq 0$ and
$\ell^2-7\ell+8<0$; so the open interval between the roots
contains no integer. In either case, $h(k,\ell)\geq 0$, and hence
$g(\ell,k,\ell)\geq 0$, for every integer
$1\leq k\leq\ell-1$. This proves $(*)$ and completes the induction.
\end{proof}

\begin{rem}
By \Cref{thm:NPhardness}, \edmonds{} cannot compute an optimal clique partition of every permutation graph, unless $\mathsf{P}=\mathsf{NP}$.  There are also small explicit examples.  Let $G$ be the permutation graph of $[2,5,4,1,7,3,6]$, depicted in \cref{fig:greedy-edmonds-counterexample}. The maximum cliques of $G$ are $\{1,4,5\}$ and $\{3,4,5\}$.  If \edmonds{} chooses $\{1,4,5\}$ in its first iteration, then the remaining graph $G-\{1,4,5\}$ has only the two edges $\{3,7\}$ and $\{6,7\}$, so a maximum matching of it has one edge, and the algorithm outputs a clique partition with $3+1=4$ edges.  However, the clique partition $\{\{3,4,5\},\{1,2\},\{6,7\}\}$ has $5$ edges.
\end{rem}

\begin{figure}[h!]
\centering
\begin{tikzpicture}[scale=.9]
    \tikzstyle{vertex}=[circle,fill=white, draw=black, inner sep=0.cm, minimum size=4.5mm]
	\begin{scope}
		\node[vertex] (2) at (-1,-2){{\tiny $2$}};
		\node[vertex] (1) at (0,-2){{\tiny $1$}};
		\node[vertex] (4) at (1,-2){{\tiny $4$}};
		\node[vertex] (3) at (2,-2){{\tiny $3$}};
		\node[vertex] (7) at (3,-2){{\tiny $7$}};
		\node[vertex] (6) at (4,-2){{\tiny $6$}};
		\node[vertex] (5) at (1,-1){{\tiny $5$}};

		\draw[-] (2) -- (1);
		\draw[-] (1) -- (4);
		\draw[-] (4) -- (3);
		\draw[-] (3) -- (7);
		\draw[-] (7) -- (6);
		\draw[-] (5) -- (1);
		\draw[-] (5) -- (3);
		\draw[-] (5) -- (4);
	\end{scope}
\end{tikzpicture}
\caption{The permutation graph of $[2,5,4,1,7,3,6]$, on which \edmonds{} can return a nonoptimal
clique partition.}
\label{fig:greedy-edmonds-counterexample}
\end{figure}

We now give a family of graphs which shows that the approximation ratio from~\Cref{edmonds1} is best possible.  Moreover, the same family shows that the improved guarantee of \Cref{edmonds1} does not carry over to \cd{} on these graphs.

\begin{thm} \label{thm:tight}
For every integer $\ell\geq 3$, there are infinitely many graphs $G$
with $\omega(G)=\ell$ and clique partitions $\mathcal{X}$ that can
be output by \edmonds{} such that
\[
   \opt=
   \frac{2\binom{\ell}{2}}{\binom{\ell}{2}+1}
   |E(\mathcal{X})|
   \qquad\text{and}\qquad
   |E(\overline{\mathcal{X}})| = 2\,\opt_{\mathrm{CD}},
\]
where $\opt$ is the maximum number of edges in a clique partition of
$G$, and $\opt_{\mathrm{CD}}=|E(G)|-\opt$ is the minimum number of edges whose deletion turns $G$ into a cluster graph.
\end{thm}

\begin{proof}
Fix $\ell\geq 3$, let
$L=\operatorname{lcm}(3,4,\ldots,\ell)$, and let $m$ be any positive
multiple of $L$. Begin with $m$ pairwise disjoint
\emph{row sets} $R_1,\ldots,R_m$, each containing $\ell$ vertices.
We recursively define disjoint sets $C_1,C_2,\ldots,C_t$. Put
$N_0=m\ell$. As long as
\[
   k_i:=\left\lceil\frac{N_{i-1}}{m}\right\rceil\geq 3,
\]
choose $C_i$ to consist of $k_i$ vertices in distinct rows and put
$N_i=N_{i-1}-k_i$. Choose these vertices from the $k_i$ largest current rows, breaking ties so that, after deleting
$C_1\cup\cdots\cup C_i$, the numbers of remaining vertices in the
rows differ by at most one (this is possible since $k_i\leq\ell
\leq m$). Let $t$ be the last index produced by this procedure.

Define $G_{\ell,m}$ on the $m\ell$ vertices by making every row
$R_j$ a clique, making every $C_i$ a clique, and adding no other
edges. See~\Cref{fig:G6} for a depiction of $G_{6,6}$.

\begin{figure} 
$$
    \begin{tikzpicture}[scale = 0.7]
    \tikzstyle{vertex}=[circle, draw=black, inner sep=0.cm, minimum size=3mm]
	\begin{scope}
		\node[vertex,fill=red] (11) at (-1,-2){};
		\node[vertex,fill=orange] (12) at (0,-2){};
		\node[vertex,fill=green] (13) at (1,-2){};
		\node[vertex,fill=green] (14) at (2,-2){};
		\node[vertex] (15) at (3,-2){};
		\node[vertex] (16) at (4,-2){};
		\draw[-] (11) -- (12) node [midway,above] {};
		\draw[-] (12) -- (13) node [midway,above] {};
		\draw[-] (13) -- (14) node [midway,above] {};
		\draw[-] (14) -- (15) node [midway,above] {};
		\draw[-] (15) -- (16) node [midway,above] {};
		\draw[-] (11) to [bend left] (16);
        \draw[-] (11) to [bend left] (15);
		\draw[-] (11) to [bend left] (14);
		\draw[-] (11) to [bend left] (13);
		\draw[-] (12) to [bend left] (16);
		\draw[-] (13) to [bend left] (16);
		\draw[-] (14) to [bend left] (16);
		\draw[-] (12) to [bend left] (14);
		\draw[-] (12) to [bend left] (15);
		\draw[-] (13) to [bend left] (15);
		\draw[-] (13) to [bend left] (16);
		\draw[-] (14) to [bend left] (16);
		\node[vertex,fill=red] (21) at (-1,-3){};
		\node[vertex,fill=orange] (22) at (0,-3){};
		\node[vertex,fill=green] (23) at (1,-3){};
		\node[vertex,fill=green] (24) at (2,-3){};
		\node[vertex] (25) at (3,-3){};
		\node[vertex] (26) at (4,-3){};
		\draw[-] (21) -- (22) node [midway,above] {};
		\draw[-] (22) -- (23) node [midway,above] {};
		\draw[-] (23) -- (24) node [midway,above] {};
		\draw[-] (24) -- (25) node [midway,above] {};
		\draw[-] (25) -- (26) node [midway,above] {};
		\draw[-] (21) to [bend left] (26);
		\draw[-] (21) to [bend left] (25);
		\draw[-] (21) to [bend left] (24);
		\draw[-] (21) to [bend left] (23);
		\draw[-] (22) to [bend left] (26);
		\draw[-] (23) to [bend left] (26);
		\draw[-] (24) to [bend left] (26);
		\draw[-] (22) to [bend left] (24);
		\draw[-] (22) to [bend left] (25);
		\draw[-] (23) to [bend left] (25);
		\draw[-] (23) to [bend left] (26);
		\draw[-] (24) to [bend left] (26);	
		\node[vertex,fill=red] (31) at (-1,-4){};
		\node[vertex,fill=orange] (32) at (0,-4){};
		\node[vertex,fill=yellow] (33) at (1,-4){};
		\node[vertex,fill=cyan] (34) at (2,-4){};
		\node[vertex] (35) at (3,-4){};
		\node[vertex] (36) at (4,-4){};
		\draw[-] (31) -- (32) node [midway,above] {};
		\draw[-] (32) -- (33) node [midway,above] {};
		\draw[-] (33) -- (34) node [midway,above] {};
		\draw[-] (34) -- (35) node [midway,above] {};
		\draw[-] (35) -- (36) node [midway,above] {};
		\draw[-] (31) to [bend left] (36);
		\draw[-] (31) to [bend left] (35);
		\draw[-] (31) to [bend left] (34);
		\draw[-] (31) to [bend left] (33);
		\draw[-] (32) to [bend left] (36);
		\draw[-] (33) to [bend left] (36);
		\draw[-] (34) to [bend left] (36);
		\draw[-] (32) to [bend left] (34);
		\draw[-] (32) to [bend left] (35);
		\draw[-] (33) to [bend left] (35);
		\draw[-] (33) to [bend left] (36);
		\draw[-] (34) to [bend left] (36);	
		\node[vertex,fill=red] (41) at (-1,-5){};
		\node[vertex,fill=orange] (42) at (0,-5){};
		\node[vertex,fill=yellow] (43) at (1,-5){};
		\node[vertex,fill=cyan] (44) at (2,-5){};
		\node[vertex] (45) at (3,-5){};
		\node[vertex] (46) at (4,-5){};
		\draw[-] (41) -- (42) node [midway,above] {};
		\draw[-] (42) -- (43) node [midway,above] {};
		\draw[-] (43) -- (44) node [midway,above] {};
		\draw[-] (44) -- (45) node [midway,above] {};
		\draw[-] (45) -- (46) node [midway,above] {};
		\draw[-] (41) to [bend left] (46);
		\draw[-] (41) to [bend left] (45);
		\draw[-] (41) to [bend left] (44);
		\draw[-] (41) to [bend left] (43);
		\draw[-] (42) to [bend left] (46);
		\draw[-] (43) to [bend left] (46);
		\draw[-] (44) to [bend left] (46);
		\draw[-] (42) to [bend left] (44);
		\draw[-] (42) to [bend left] (45);
		\draw[-] (43) to [bend left] (45);
		\draw[-] (43) to [bend left] (46);
		\draw[-] (44) to [bend left] (46);	
		\node[vertex,fill=red] (51) at (-1,-6){};
		\node[vertex,fill=orange] (52) at (0,-6){};
		\node[vertex,fill=yellow] (53) at (1,-6){};
		\node[vertex,fill=cyan] (54) at (2,-6){};
		\node[vertex,fill=purple] (55) at (3,-6){};
		\node[vertex] (56) at (4,-6){};
		\draw[-] (51) -- (52) node [midway,above] {};
		\draw[-] (52) -- (53) node [midway,above] {};
		\draw[-] (53) -- (54) node [midway,above] {};
		\draw[-] (54) -- (55) node [midway,above] {};
		\draw[-] (55) -- (56) node [midway,above] {};
		\draw[-] (51) to [bend left] (56);
		\draw[-] (51) to [bend left] (55);
		\draw[-] (51) to [bend left] (54);
		\draw[-] (51) to [bend left] (53);
		\draw[-] (52) to [bend left] (56);
		\draw[-] (53) to [bend left] (56);
		\draw[-] (54) to [bend left] (56);
		\draw[-] (52) to [bend left] (54);
		\draw[-] (52) to [bend left] (55);
		\draw[-] (53) to [bend left] (55);
		\draw[-] (53) to [bend left] (56);
		\draw[-] (54) to [bend left] (56);	
		\node[vertex,fill=red] (61) at (-1,-7){};
		\node[vertex,fill=yellow] (62) at (0,-7){};
		\node[vertex,fill=yellow] (63) at (1,-7){};
		\node[vertex,fill=purple] (64) at (2,-7){};
		\node[vertex,fill=purple] (65) at (3,-7){};
		\node[vertex] (66) at (4,-7){};
		\draw[-] (61) -- (62) node [midway,above] {};
		\draw[-] (62) -- (63) node [midway,above] {};
		\draw[-] (63) -- (64) node [midway,above] {};
		\draw[-] (64) -- (65) node [midway,above] {};
		\draw[-] (65) -- (66) node [midway,above] {};
		\draw[-] (61) to [bend left] (66);
		\draw[-] (61) to [bend left] (65);
		\draw[-] (61) to [bend left] (64);
		\draw[-] (61) to [bend left] (63);
		\draw[-] (62) to [bend left] (66);
		\draw[-] (63) to [bend left] (66);
		\draw[-] (64) to [bend left] (66);
		\draw[-] (62) to [bend left] (64);
		\draw[-] (62) to [bend left] (65);
		\draw[-] (63) to [bend left] (65);
		\draw[-] (63) to [bend left] (66);
		\draw[-] (64) to [bend left] (66);	
\end{scope}     
\end{tikzpicture}
$$
\caption{The graph $G_{6,6}$, with $k_1=6,k_2=5,k_3=5,k_4=4,k_5=3,k_6=3$. Vertices of the same (non-white) color form a clique in the graph.}
\label{fig:G6}
\end{figure}

We first record the clique structure of this graph. If a
clique contains vertices from two different rows, then those two
vertices must lie in a common $C_i$. Since the sets $C_i$ are
disjoint and each $C_i$ meets each row in at most one vertex, every
other vertex of the clique must also lie in that same $C_i$.
Consequently, every clique of $G_{\ell,m}$ is contained in a row or
in one of the sets $C_i$. In particular,
\[
   \omega(G_{\ell,m})=\ell.
\]

After $C_1,\ldots,C_{i-1}$ have been deleted, the remaining row
sizes differ by at most one and have sum $N_{i-1}$. Their maximum
size is therefore $\lceil N_{i-1}/m\rceil=k_i$. The future sets
$C_j$ have nonincreasing sizes $k_j\leq k_i$, and the preceding
clique-structure observation remains valid in the induced graph.
Thus, $C_i$ is a maximum clique at iteration $i$, and \edmonds{} can
choose $C_1,C_2,\ldots,C_t$ in that order.

At termination, $N_t\leq 2m$. Just before the last deletion there
were more than $2m$ vertices, and $k_t\leq\ell\leq m$, so $N_t>m$.
The remaining row sizes are therefore all $1$ or $2$. There are
exactly $N_t-m$ rows of size $2$, and a maximum matching in the
remaining graph has $N_t-m$ edges. Hence this run of \edmonds{}
outputs a partition $\mathcal{X}_{\ell,m}$ with
\begin{equation} \label{eq:tight-output}
   |E(\mathcal{X}_{\ell,m})|
   =\sum_{i=1}^t\binom{k_i}{2}+N_t-m.
\end{equation}

The row partition shows that
$\opt_{\ell,m}\geq m\binom{\ell}{2}$. The reverse inequality follows
from $\omega(G_{\ell,m})=\ell$. If a clique partition has part sizes
$s_1,s_2,\ldots$, then
\[
   \sum_j\binom{s_j}{2}
   \leq\frac{\ell-1}{2}\sum_j s_j
   =\frac{\ell-1}{2}m\ell
   =m\binom{\ell}{2}.
\]
Therefore,
\begin{equation} \label{eq:tight-optimum}
   \opt_{\ell,m}=m\binom{\ell}{2}.
\end{equation}

It remains to evaluate \eqref{eq:tight-output}. This can be done
exactly because $m$ is divisible by every
$q\in\{3,\ldots,\ell\}$. Initially $N_0=\ell m$. While
$N_{i-1}$ decreases from $qm$ to $(q-1)m$, we have
$k_i=\lceil N_{i-1}/m\rceil=q$. This phase consists of exactly
$m/q$ steps, each deleting $q$ vertices. Inducting downwards from
$q=\ell$ to $q=3$ shows that $k_i=q$ occurs exactly $m/q$ times and
that $N_t=2m$. Therefore, \eqref{eq:tight-output} gives
\begin{align*}
|E(\mathcal{X}_{\ell,m})|
 &=\sum_{q=3}^{\ell}\frac{m}{q}\binom{q}{2}+m\\
 &=\frac{m}{2}\sum_{q=3}^{\ell}(q-1)+m\\
 &=\frac{m}{2}\left(\binom{\ell}{2}+1\right).
\end{align*}
Together with \eqref{eq:tight-optimum}, this yields the exact equality
\[
\opt_{\ell,m}
=\frac{2\binom{\ell}{2}}{\binom{\ell}{2}+1}
 |E(\mathcal{X}_{\ell,m})|.
\]

Finally, we count the deleted edges. The edge set of $G_{\ell,m}$ is the disjoint union of the edge sets of the $m$ row cliques and of the cliques $C_1,\ldots,C_t$, so by the computation above,
\[
   |E(G_{\ell,m})|
   = m\binom{\ell}{2}+\sum_{i=1}^{t}\binom{k_i}{2}
   = m\binom{\ell}{2}+\frac{m}{2}\left(\binom{\ell}{2}-1\right).
\]
Hence, by \eqref{eq:tight-optimum}, the minimum number of edges whose deletion turns $G_{\ell,m}$ into a cluster graph is
\[
   \opt_{\mathrm{CD}}=|E(G_{\ell,m})|-\opt_{\ell,m}
   =\frac{m}{2}\left(\binom{\ell}{2}-1\right),
\]
whereas the run of \edmonds{} above deletes
\[
   |E(\overline{\mathcal{X}_{\ell,m}})|
   =|E(G_{\ell,m})|-|E(\mathcal{X}_{\ell,m})|
   =m\binom{\ell}{2}-m
   =2\,\opt_{\mathrm{CD}}
\]
edges. There are infinitely many positive multiples $m$ of $L$, so this
gives infinitely many examples.
\end{proof}

Since every run of \edmonds{} is also a run of \greedy{}, \Cref{thm:tight} also shows that the factor $2$ in \Cref{thm:greedy2} is best possible for \greedy{} on graphs of any fixed clique number $\ell\geq 3$.  

\appendix
\section{Proofs of the greedy approximation theorems} \label{app:greedy}
\begin{proof}[Proof of~\Cref{thm:greedy1}]
    Let $\mathcal{X}:=(X_1, \dots, X_\ell)$ be the sequence of cliques constructed by \greedy{}, where $X_1$ is a maximum clique of $G$.  We proceed by induction on $\ell$.  If $\ell=1$, then $|E(\mathcal{X})| = \opt$.  Thus, we may assume $\ell \geq 2$.  Let $\mathcal{Y}:=(Y_1, \dots, Y_m)$ be an optimal clique partition of $G$.  
    Let $G':=G-X_1$, $\mathcal{X}':=(X_2, \dots, X_\ell)$, $\mathcal{Y}':=(Y_1 \setminus X_1, \dots, Y_m \setminus X_1)$, and $\opt'$ be the number of edges in an optimal clique partition of $G'$.  Let $k:=|X_1|$.  For each vertex $x$ in $X_1$, let $\mathcal{Y}(x)$ be the unique cluster $Y$ of $\mathcal{Y}$ such that $x \in Y$.  Define $y(x):=|\mathcal{Y}(x)|-1$.  
    Since $X_1$ is a maximum size clique of $G$, $y(x) \leq k-1$ for all $x \in X_1$.   
    Therefore, 
 \[
    |E(\mathcal{Y})|-|E(\mathcal{Y}')|
    \leq \sum_{x \in X_1} y(x)
    \leq \sum_{x \in X_1}(k-1)=k(k-1).
    \]
  
    We also clearly have
      \[
    |E(\mathcal{X})|-|E(\mathcal{X}')| =\frac{k(k-1)}{2}.
    \]
    By induction, $|E(\mathcal{X}')| \geq \frac{\opt'}{2} \geq \frac{|E(\mathcal{Y}')|}{2}$. Since
    $|E(\mathcal{Y})|\leq |E(\mathcal{Y}')|+k(k-1)$, the preceding
    two inequalities give
    $|E(\mathcal{X})| \geq \frac{|E(\mathcal{Y})|}{2}$, as required.
\end{proof}

\begin{proof}[Proof of~\Cref{thm:greedy2}]
Let $\mathcal{X}=(X_1,\ldots,X_r)$ be the sequence of cliques
constructed by \greedy{}, where $X_1$ is a maximum clique of $G$.
We induct on $r$. The case $r=1$ is immediate. Let
$\mathcal{Y}=(Y_1,\ldots,Y_m)$ be an optimal clique partition, and
put
\[
\begin{split}
G'&=G-X_1,\qquad
\mathcal{X}'=(X_2,\ldots,X_r),\\
\mathcal{Y}'&=(Y_1\setminus X_1,\ldots,Y_m\setminus X_1),\\
\mathcal{Z}&=(X_1,Y_1\setminus X_1,\ldots,Y_m\setminus X_1),
\end{split}
\]
omitting empty parts. Write $D(\mathcal{P})=E(\overline{\mathcal{P}})$
for any clique partition $\mathcal{P}$. By induction,
\[
   |D(\mathcal{X}')|\leq 2|D(\mathcal{Y}')|.
\]
Since $D(\mathcal{Z})$ is the disjoint union of
$\delta_G(X_1)$ and $D(\mathcal{Y}')$, it follows that
\begin{equation} \label{eq:greedy-cd-step}
|D(\mathcal{X})|
\leq |\delta_G(X_1)|+2|D(\mathcal{Y}')|
=|D(\mathcal{Z})|+|D(\mathcal{Y}')|.
\end{equation}

Let
\[
   P=D(\mathcal{Z})\setminus D(\mathcal{Y})
   \quad\text{and}\quad
   Q=D(\mathcal{Y})\setminus D(\mathcal{Z}).
\]
Put $k=|X_1|$ and $x_i=|X_1\cap Y_i|$. Because every $Y_i$ is a
clique and $X_1$ is a maximum clique, $|Y_i|\leq k$, and hence
\begin{align*}
|P|
 &=\sum_{i=1}^m x_i(|Y_i|-x_i)\\
 &\leq\sum_{i=1}^m x_i(k-x_i)
 =2\sum_{1\leq i<j\leq m}x_ix_j
 =2|Q|.
\end{align*}
Here $Q$ consists precisely of the edges of the clique $X_1$ whose
ends lie in different parts of $\mathcal{Y}$. Moreover,
$D(\mathcal{Y}')$ and $Q$ are edge-disjoint subsets of
$D(\mathcal{Y})$: the former contains edges with both ends outside
$X_1$, while the latter contains edges with both ends inside $X_1$.
Using \eqref{eq:greedy-cd-step}, we obtain
\begin{align*}
|D(\mathcal{X})|
&\leq |D(\mathcal{Z})|+|D(\mathcal{Y}')|\\
&=|D(\mathcal{Y})|+|P|-|Q|+|D(\mathcal{Y}')|\\
&\leq |D(\mathcal{Y})|+|Q|+|D(\mathcal{Y}')|\\
&\leq 2|D(\mathcal{Y})|=2\opt. \qedhere
\end{align*}
\end{proof}

\section*{Funding}
Tony Huynh is supported by the Institute for Basic Science (IBS-R029-C1).  Nicola Galesi was partly supported by PRIN 2022 project  "Logical methods in combinatorics", 2022BXH4R5, MIUR
(Italian Ministry of University and Research).



\section*{AI Declaration}
The main ideas of the proof of~\cref{thm:NPhardness} were found by ChatGPT 5.6 Sol. We remark that the ChatGPT proof of~\cref{thm:NPhardness} contained some serious errors, and has been completely rewritten by the authors.  The example in \Cref{prop:lp-perm} was found by a computer search carried out with the assistance of Anthropic Claude.  All other results in this paper were obtained by humans.  OpenAI Codex and Anthropic Claude were also used for
language editing, proof checking, and to search for additional references. After using
these tools, the authors reviewed and edited the content as needed and
take full responsibility for the content of the article.

\bibliographystyle{cas-model2-names}
\bibliography{references}
\end{document}